\documentclass[aps,prl,reprint]{revtex4-1}

\usepackage{gensymb}
\usepackage{amsmath}
\usepackage{amssymb}
\usepackage{bm}
\usepackage{color}
\usepackage[linkcolor=blue, colorlinks=true, urlcolor=blue,
citecolor=red]{hyperref}
\usepackage{graphicx}
\usepackage[caption=false]{subfig}
\usepackage{bbold}
\usepackage{verbatim}
\usepackage{appendix}
\usepackage{amsthm}
\usepackage[normalem]{ulem}

\theoremstyle{plain}

\newcommand{\ket}[1]{\left| #1 \right>} 
\newcommand{\bra}[1]{\left< #1 \right|} 
\newcommand{\braket}[2]{\left\langle #1|#2 \right\rangle}
\newcommand{\ketbra}[1]{\ket{#1}\hspace*{-0.9mm}\bra{#1}}
\newcommand{\ketbraOD}[2]{\ket{#1}\hspace*{-0.9mm}\bra{#2}}

\newcommand{\Eq}[1]{Eq.~\eqref{#1}}
\newcommand{\av}[1]{\left\langle #1 \right\rangle}
\newcommand{\Id}{\mathbb{1}}
\newcommand{\Tr}[1]{\text{Tr}\left[#1\right]} 
\newcommand{\TrS}[2]{\text{Tr}_{#1}\left[#2\right]}
\newcommand{\aop}{\hat{a}}
\newcommand{\bop}{\hat{b}}

\begin{document}

\title{Nonclassicality Criteria in Multiport Interferometry}

\author{L. Rigovacca$^1$}
\email{l.rigovacca14@imperial.ac.uk}
\author{C. Di Franco$^{1,2,3}$}
\author{B. J. Metcalf$^4$}
\author{I. A. Walmsley$^4$}
\author{M. S. Kim$^1$}

\affiliation{$^1$Blackett Laboratory, Imperial College London, London, SW7 2AZ, United Kingdom}
\affiliation{$^2$School of Physical and Mathematical Sciences, Nanyang Technological University, 637371, Singapore}
\affiliation{$^3$Complexity Institute, Nanyang Technological University, 637723, Singapore}
\affiliation{$^4$Clarendon Laboratory, University of Oxford, Parks Road, Oxford, OX1 3PU, United Kingdom}

\begin{abstract}
Interference lies at the heart of the behavior of classical and quantum light. It is thus crucial to understand the boundaries between which interference patterns can be explained by a classical electromagnetic description of light and which, on the other hand, can only be understood with a proper quantum mechanical approach. While the case of two-mode interference has received a lot of attention, the multimode case has not yet been fully explored. Here we study a general scenario of intensity interferometry: we derive a bound on the average correlations between pairs of output intensities for the classical wavelike model of light, and we show how it can be violated in a quantum framework. As a consequence, this violation acts as a nonclassicality witness, able to detect the presence of sources with sub-Poissonian photon-number statistics. 
We also develop a criterion that can certify the impossibility of dividing a given interferometer into two independent subblocks.
\end{abstract}
%
%

\maketitle

Hong, Ou, and Mandel (HOM) discovered that if two independent and indistinguishable photons, in pure quantum states, impinge on the two input ports of a balanced beam splitter, they always bunch together and exit the apparatus from the same output port \cite{HOM_PRL_1987}.
This simple effect has many consequences, e.g., in distinguishability testing \cite{Sun2009}, linear-optical quantum computing \cite{KLM_LOcomputing}, entanglement detection \cite{Shih1988} or swapping \cite{Zbinden2007}, and metrology \cite{n00n_1,n00n_2,Giovannetti_MetrologyReview,DemkowiczDobrzański2015345}.
The nonclassicality of this phenomenon can be well understood by repeating the experiment many times, and by recording the intensities $I_1$, $I_2$ at the two output ports: labeling the average over many runs by $\av{\cdot}$, the correlation function
\begin{equation}\label{def: G_12}
G_{12} = \frac{\av{I_1 I_2}}{\av{I_1}\av{I_2}}
\end{equation}
will be zero in the ideal case, because on each run the intensity at one of the two ports will vanish.
$G_{12}$ has a well-defined classical limit, which makes it a suitable candidate to use in distinguishing quantum light beams from classical ones. We can either consider completely distinguishable photons, i.e., single excitations occupying orthogonal space-time modes, or pulses of classical light, described by electromagnetic fields. In both cases, if they are emitted by statistically independent sources and injected into the beam splitter, $G_{12}$ is constrained to be greater than or equal to $1/2$ \cite{Mandel_ClBound,Paul_ClBound}. Therefore, the value $G_{12}=0$ obtained in the ideal HOM effect represents a strong signature of nonclassicality.

Several authors have investigated interference effects of noninteracting particles with the aim of reproducing or generalizing HOM's result to different situations (see Refs. \cite{Tichy2010,Lim2005}) not necessarily constrained to linear optics \cite{HOM_atoms,HOM_phonons,Jachura:15,Sokolovski2016,Urbina2016}. 
However, photonics remains the physical platform of choice for these studies, since it is now possible to prepare and manipulate several photons in ambient laboratory conditions, which can then be injected into multimode interferometers \cite{2016arXiv160306984S,Tillmann2015,Rome_experiment,Oxford_experiment,Crespi2013,Spring2013,Yao2012,Matthews2011}. The recent investigations of many-particle interference effects have revealed a need for a deeper understanding of the phenomenon. On the computational side, boson sampling is a feasible candidate to show the possibility of outperforming classical computers exploiting the laws of quantum mechanics
\cite{Aaronson2011,Shchesnovich2016,Bentivegna2015,Broome2013,Crespi2013,Spring2013,Spagnolo2014}. From a foundational perspective, on the other hand, the interplay between the wavelike behavior of photons and the many-particle interference effects arising due to their bosonic nature is not well understood \cite{Tichy2012,Belinskii1992,Lim2005,Mayer2011,Laibacher2015,Tichy2011,Ra2013}. This is an important issue because these two features heavily influence the probabilities of detection events, often leading to counterintuitive results \cite{Tichy2012,Tichy2011,Ra2013}.
Typically these studies compare their findings with the evolution of completely distinguishable photons: this retains the quantization of the number of particles, but removes interparticle interference.

In this Letter we make the complementary choice, by
studying the alternative classical regime where
independent sources emit light pulses fully described in terms of their electric field.   
As already mentioned, in a situation with two sources and two detectors, the classical bound $G_{12}\geq 1/2$ holds, which can be maximally violated by the HOM setup, i.e., $G_{12}=0$. It is, therefore, natural to ask how this result can be extended to a more general framework, with an arbitrary number of sources and detectors. Here we provide an answer to this question, by finding a tight lower bound for the correlations that can arise among the output intensities of a generic multiport interferometer, when the aforementioned classical sources are used. By using a quantum mechanical approach, we then study ``if'' and ``by how much'' quantum input states of light can violate this threshold. Finally, we show how our findings allow us to develop a sufficient criterion that can certify the impossibility of dividing an interferometer into independent subblocks.

\emph{Correlation function.---} In characterizing the correlations among several detected intensities, say, $M\geq2$, we have to choose a generalization of $G_{12}$, defined in Eq.~\eqref{def: G_12}. Recently, several authors have considered higher-order correlation functions (i.e., in which the average involves products of more than two intensities) to study multiparticle interference \cite{Mayer2011,Tamma2015} or to obtain advantages in imaging resolution \cite{HighOrder_1,HighOrder_2,HighOrder_3,HighOrder_4}.
On the other hand, Walschaers \emph{et al.} showed that the simpler quantity $\av{I_i I_j} - \av{I_i}\av{I_j}$ can yield information on the statistics of the interfering particles \cite{Walschaers2016BS} and on their distinguishability \cite{Walschaers2016} if it is averaged over many output modes of a generic interferometer.
We introduce a similar but slightly different quantifier, obtained as the \emph{normalized} average,
\begin{equation}\label{def: Gbar}
\overline{G}  = \frac{1}{\binom{M}{2}} \sum_{i < j} \frac{\av{I_i I_j}}{\av{I_i}\av{I_j}},
\end{equation}
where $i < j$ enforces the sum to be over all pairs of detectors. The normalization chosen in \Eq{def: Gbar} assures that $\overline{G}$ does not depend on the average intensities, but only on their correlations: this is a necessary condition to obtain a classical bound independent of the total intensities of the sources.
With respect to considering higher-order correlators, $\overline{G}$ has the advantage of being composed of many simpler contributions, while still taking into account all available data. Moreover, the measure of $\overline G$ only requires the simultaneous intensity readouts in two detectors, thus allowing a quicker experimental estimation in the presence of detector inefficiencies. We also point out that the experimental effort required to estimate $\av{I_i I_j}$, and therefore $\overline{G}$, has a polynomial scaling with the number of sources and photon-number resolving detectors \cite{Walschaers2016BS}.
Finally, note that only detectors with $\av{I_i}\neq 0$ (i.e., receiving a nonzero amount of light) should be considered in the average of \Eq{def: Gbar}, in order to keep $\overline G$ well defined.

\begin{figure}
	\centering
	\includegraphics[scale=0.9]{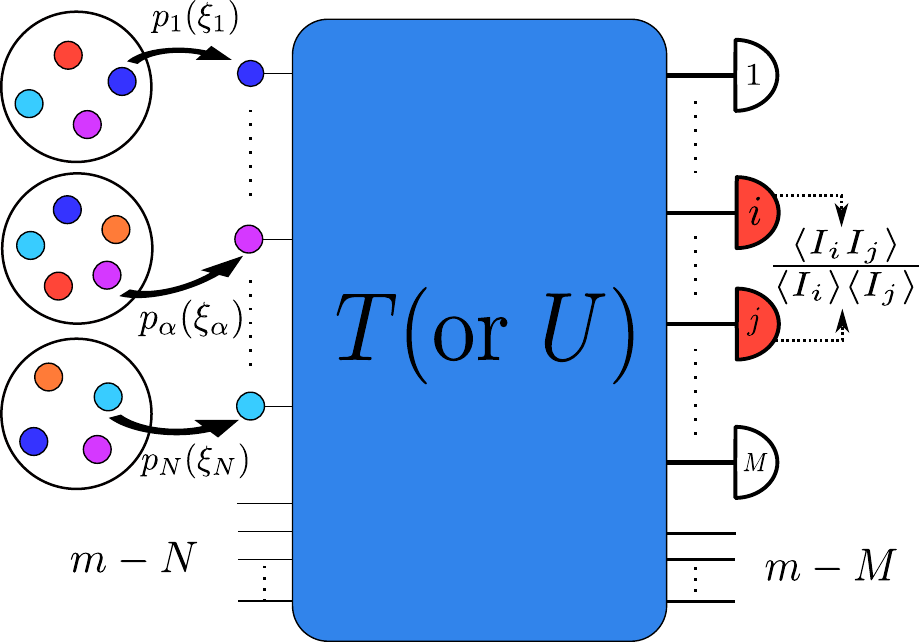}
	\caption{Sketch of the setup described in the main text, with $N$ sources and $M$ detectors. In the classical (quantum) picture the interferometer is characterized by the transfer matrix $T$ ($U$), while the colored circles represent the light pulses $\vec{E}_{\alpha,\xi_\alpha}(t)$ (the states $\big|\varphi_\alpha^{(\xi_\alpha)}\big\rangle$). In the quantum framework the number $m$ of interferometric modes has to be greater than $N$ and $M$: if the inequalities are strict there will be vacuum inputs and nonmonitored outputs. If at least two detections are successful the event can be used for the evaluation of $\overline G$. \label{figure}}
\end{figure}

\emph{Description of the setup.---}
We now describe the interferometric setup, sketched in Fig.~\ref{figure}, in the classical and quantum  scenario. In both cases pulses of light are emitted by $N$ sources and are detected by $M$ detectors after a linear evolution. As in the HOM setup, in each realization the phases of the pulses are chosen randomly in $[0,2\pi]$, the sources are independent, and are allowed to be stochastic.
More precisely, at every run of the experiment in the classical case the $\alpha$th source emits the electric field $\vec{E}_{\alpha,\xi_\alpha}(t) = A_\alpha(\xi_\alpha) \vec\zeta^{(\xi_\alpha)}(t)$ with probability $p_\alpha(\xi_\alpha)$. Here $A_\alpha(\xi_\alpha)$ is a complex number whose phase changes randomly from one pulse to the other, while $\vec\zeta_\alpha^{(\xi_\alpha)}(t)$ defines the mode of the field. In particular, if $\{\vec{\epsilon}_{\omega,\lambda}\}_\lambda$ are orthonormal polarization vectors, one has
\begin{equation}\label{def: electric field}
\vec\zeta_\alpha^{(\xi_\alpha)}(t)= \sum_\lambda \int \textit{d}\omega\, \sqrt{\frac{\hbar \omega}{2\pi}} \, g_{\omega,\lambda}(\xi_\alpha) \,e^{-i \omega t} \vec{\epsilon}_{\omega,\lambda},
\end{equation}
where the coefficients  $\{g_{\omega,\lambda}(\xi_\alpha)\}$ depend on the value of the random variable $\xi_\alpha$ and satisfy the relation $\sum_{\lambda} \int \textit{d}\omega\,|g_{\omega,\lambda}(\xi_\alpha)|^2=1$. We added the factor $\hbar \omega$ only to ease the comparison with the quantum case later on, but classically it could be included in $g_{\omega,\lambda}$ and $A_\alpha$.
The linear evolution can then be represented via a complex transfer matrix $T$, which maps the input fields $\{\vec{E}_{\alpha,\xi_\alpha}\}$ to those at the detectors' positions, labeled by $i = 1,\ldots, M$:
\begin{equation}\label{eq: I/O relation}
\vec{O}_{i,\vec{\xi}}(t) = \sum_{\alpha=1}^{N} T_{i\alpha} \vec{E}_{\alpha,\xi_\alpha}(t-\tau_{i\alpha}),
\end{equation}
where $\tau_{i\alpha}$ are the evolution times and $\vec{\xi} = (\xi_1,\ldots,\xi_N)$. The intensity measured by the $i$th detector will therefore be, up to a dimensional proportionality factor,
\begin{equation}\label{eq: output intensities}
I_{i,\vec{\xi}} = \int\limits_{-\tau_M/2}^{+\tau_M/2} \text{d} t \; \vec{O}^*_{i,\vec{\xi}}(t) \cdot \vec{O}_{i,\vec{\xi}}(t),
\end{equation}
where we consider the measurement time $\tau_M$ to be much longer than the other time scales, so that the integration can be equivalently performed over the whole real axis. This means that the detectors measure a quantity proportional to the energy, or integrated flux, of the light pulses. The quantities $\av{I_i}$ and $\av{I_i I_j}$ are then obtained by averaging over the configurations $\{\vec{\xi}\}$.
Note that their evaluation is simplified by the following relations imposed by the randomness of the phase characterizing each pulse of light: $\big\langle A_\alpha\big\rangle = \big\langle A_\alpha A_\beta\big\rangle = 0$ and $\big\langle A^*_\alpha A_\beta\big\rangle = \delta_{\alpha,\beta} \av{|A_\alpha|^2}$.

In order to move to a quantum picture, an operator for the electric field emitted by each source is defined as
\begin{equation}
\hat E_\alpha(t) = \sum_{\lambda} \int \textit{d}\omega \, \sqrt{\frac{\hbar \omega}{2\pi}} \,  \aop_{\alpha;\omega,\lambda} \, e^{-i\omega t} \vec{\epsilon}_{\omega,\lambda},
\end{equation}
up to an irrelevant factor, where the bosonic annihilation operators $\{\aop_{\alpha;\omega,\lambda}\}$ satisfy the canonical commutation relations $[\aop_{\alpha;\omega,\lambda},\aop_{\beta;\omega^\prime,\lambda^\prime}^\dagger] = \delta_{\alpha,\beta} \delta_{\lambda,\lambda^\prime}\delta(\omega-\omega^\prime)$ (see Ref. \cite{Loudon_1990}).
With probability $p_\alpha(\xi_\alpha)$, the $\alpha$th source then emits the quantum state
\begin{equation}\label{eq: quantum state}
\big|\varphi_\alpha^{(\xi_\alpha)}\big\rangle = \sum_{n=0}^\infty \varphi_\alpha^{(\xi_\alpha)}(n) \left(\aop^\dagger_{\alpha;\xi_\alpha}\right)^n \ket{0},
\end{equation}
where the amplitudes $\{\varphi_\alpha^{(\xi_\alpha)}(n)\}_n$ depend on the random variable $\xi_\alpha$, $\ket{0}$ represents the vacuum, and $\aop_{\alpha;\xi_\alpha}$ can be decomposed as \begin{equation}\label{eq: operator}
\aop_{\alpha;\xi_\alpha} = \sum_{\lambda} \int \textit{d}\omega \, g_{\omega,\lambda}^*(\xi_\alpha) \, \aop_{\alpha;\omega,\lambda}.
\end{equation} 
However, note that the coherence between different photon numbers in \Eq{eq: quantum state} is lost once we average over many realizations of the same pulse, because each of them is emitted with a random phase multiplying $g^*_{\omega,\lambda}$ in \Eq{eq: operator}. In this scenario, it is natural to inject the emitted states in an $m\times m$ linear optical interferometer, with $m\geq M,N$. For any given realization $\vec{\xi}$ of the sources, the output modes will be characterized by the set of operators $\{\hat b_{i;\vec{\xi}}\}_{i=1}^m$, obtained as $\hat b_{i;\vec{\xi}} = \sum_{\alpha=1}^{m} U_{i\alpha} \, \hat a_{\alpha;\xi_\alpha}$, where $U$ is a unitary matrix that plays a role analogous to the classical $T$. 
The intensities appearing in \Eq{def: Gbar} are obtained, up to a dimensional proportionality factor, by taking the expectation value of the operators $\hat I_i$ and $\hat I_i \hat I_j$ on the input state, where 
\begin{equation}
	\hat{I}_i = \int\limits_{-\tau_M/2}^{+\tau_M/2} \textit{d}t \, \hat{E}^\dagger_{i}(t)\hat{E}_{i}(t) = \sum_{\lambda}\int \textit{d}\omega \, \bop_{i;\omega,\lambda}^\dagger \bop_{i;\omega,\lambda} \,\hbar \omega.
\end{equation}

For the sake of simplicity, in the following we will assume that the mode of the emitted fields is characterized by the same weights $g_{\omega,\lambda}(\xi_0)$ for all sources and all realizations, and that the evolution times $\tau_{i\alpha}$ are all the same. Intuitively, these conditions maximize the interference and in the classical case lead to the minimum value of \Eq{def: Gbar} (see Ref. \cite{SM} for the proof). With these hypotheses, it turns out that the averaged intensities that appear in the classical or quantum expression of $\overline G$ are proportional to
$\mathcal E = \sum_{\lambda}\int \textit{d}\omega \, \hbar \omega |g_{\omega,\lambda}(\xi_0)|^2$ \cite{SM}, which represents the energy associated with the chosen mode. In particular, classically one has $I_{i,\vec{\xi}} = \mathcal E A_i(\vec\xi)^*A_i(\vec\xi)$, where $A_i(\vec\xi) = \sum_{\alpha=1}^{N} T_{i_\alpha} A_\alpha(\xi_\alpha)$, while in the quantum case, one finds 
\begin{align}
\big\langle\hat I_i\big\rangle &= \mathcal E  \,\text{Tr}\big[\hat\rho \,\bop_{i;\xi_0}^\dagger \bop_{i;\xi_0}\big], \label{eq: quant Ii}\\
\big\langle\hat I_i \hat I_j\big\rangle &= \mathcal E^2\,\text{Tr}\big[\hat \rho \,\bop_{i;\xi_0}^\dagger \bop_{j;\xi_0}^\dagger  \bop_{i;\xi_0} \bop_{j;\xi_0}\big].
\label{eq: quant IiIj}
\end{align}
This is intuitive because the intensity of the quantum field is directly connected with the photon number, when each photon carries the same amount of energy.
In Eqs. \eqref{eq: quant Ii} and \eqref{eq: quant IiIj}, $\hat \rho$ is the average emitted state
\begin{equation}\label{def: rho}
\hat\rho = \bigotimes_{\alpha=1}^N \sum_{n_\alpha} \frac{q_\alpha(n_\alpha)}{n_\alpha!} \,(\aop_{\alpha,\xi_0}^{\dagger})^ {n_\alpha} \ketbra{0}(\aop_{\alpha,\xi_0})^{n_\alpha},
\end{equation}
where $q(\vec{n}) = q_1(n_1)\ldots q_N(n_N)$ is the effective probability distribution of the process \cite{footnote1}.
In the following, when \Eq{def: Gbar} is calculated in the classical or quantum regime, it will be written, respectively, as $\overline G^{(cl)}$ or $\overline G^{(Q)}$. We will also drop the label $\xi_0$ from the bosonic operators.

\emph{Classical bound.---} We now look for the minimum value that $\overline G^{(cl)}$ can take. This will be the benchmark against which the results of an experiment must be compared in order to certify a nonclassical behavior, i.e., the impossibility of simulating the same result with \emph{only} classical resources. Explicit calculations yield $\av{I_i} = \mathcal E \sum_{\alpha = 1}^{N} |T_{i\alpha}|^2 \av{|A_\alpha|^2}$ and
\begin{align}\label{eq: <Ii Ij>}
\av{I_i I_j} &= \av{I_i}\av{I_j} + \mathcal E^2 \sum_{\alpha\neq \beta}^{N} T_{i\alpha}T_{i\beta}^* T_{j\beta}T_{j\alpha}^* \av{|A_\alpha|^2} \av{|A_\beta|^2} \notag\\
&+ \mathcal E^2 \sum_{\alpha = 1}^{N} |T_{i\alpha}|^2 |T_{j\alpha}|^2 \left[\av{|A_\alpha|^4} - \av{|A_\alpha|^2}^2\right],
\end{align}
where the last (positive) term vanishes for sources with fixed intensity, which are, therefore, optimal. The remaining minimization can be performed by defining a set of $M$ normalized vectors $\psi_i
\in \mathbb{C}^N$ with components $\psi_i(\alpha)= T_{i\alpha}^* \sqrt{\mathcal E \av{|A_\alpha|^2}/\av{I_i}}$,
which allow us to rewrite $\overline G^{(cl)}$ as
\begin{equation}\label{eq: class expr}
\overline G^{(cl)} = 1 + \frac{1}{\binom{M}{2}}\sum_{i< j}^M \Big[|\psi_i^* \cdot \psi_j|^2 - \sum_{\alpha=1}^N |\psi_i(\alpha)\psi_j(\alpha)|^2\Big],
\end{equation}
where $\psi_i^* \cdot \psi_j = \sum_{\alpha}\psi_i^*(\alpha) \cdot \psi_j(\alpha)$.
This expression can be minimized with respect to the vectors $\psi_i$ (see the Appendix), yielding
\begin{equation}\label{res: classical bound}
\min \overline{G}^{(cl)}_{N,M} =  \begin{cases}
 1-\frac{N-1}{N(M-1)} & \text{ if } N\leq M\\
 1 - \frac{1}{M}  & \text{ if } N \geq M,
 \end{cases}
\end{equation}
where the subscripts on the left-hand side emphasize the dependence on the number of sources and detectors (respectively, $N$ and $M$).
We can verify that the minimum is reached by letting  light fields with the same input intensity evolve with a highly symmetric $T$: the $M\times M$ Fourier transform matrix (FTM), whose $(j,\alpha)$ element is given by $ e^{2\pi i (j-1)(\alpha-1)/M}/\sqrt{M}$. Intuitively, this interferometric apparatus yields the minimum output correlations because it leads to a high degree of interference: the input intensities are equally split among all outputs and the phases are symmetrically distributed over $2\pi$.
We point out that, when $N > M$, the configuration that achieves the $1-1/M$ bound completely ignores $N-M$ sources (see the Appendix), whose light fields never reach the detectors. This setup is, therefore, effectively equivalent to a symmetric one smaller in size, with only $M$ sources and detectors.

\emph{Quantum description.---} The explicit evaluation of Eqs. \eqref{eq: quant Ii} and \eqref{eq: quant IiIj} follows the steps of other studies (see Refs. \cite{Mayer2011,Walschaers2016BS,Walschaers2016}), and for the input state in \Eq{def: rho}, one finds
$\big\langle\hat I_i\big\rangle = \mathcal E \sum_{\alpha = 1}^{m} |U_{i\alpha}|^2 \av{\hat n_\alpha}_q$ and
\begin{align}\label{eq: quantum <I1 Ij>}
&\big\langle\hat I_i \hat  I_j\big\rangle = \big\langle\hat I_i\big\rangle \big\langle\hat I_j\big\rangle +  \mathcal E^2
\sum_{\alpha\neq \beta}^{m} U_{i\alpha}U_{i\beta}^* U_{j\beta}U_{j\alpha}^* \; \av{\hat n_\alpha}_q \av{\hat n_\beta}_q \notag \\
& \quad+ \mathcal E^2 \sum_{\alpha=1}^{m} |U_{i\alpha}|^2 |U_{j\alpha}|^2 \left[\left(\av{\hat n_\alpha^2}_q - \av{\hat n_\alpha}^2_q\right) - \av{\hat n_\alpha}_q\right],
\end{align}
where $\hat n_\alpha = \aop_\alpha^\dagger \aop_\alpha$ and the subscript $q$ reminds us of the effective probability distribution appearing in \Eq{def: rho}. Note that if there are more interferometric modes than sources (i.e., $N < m$), we can trivially extend the definition of $\hat \rho$ in \Eq{def: rho} by considering $q_{\alpha>N}(n_\alpha)=\delta_{n_\alpha,0}$.
Apart from the natural correspondences $U \leftrightarrow T$ and $\av{\hat n_\alpha}_q \leftrightarrow \av{|A_\alpha|^2}$, we can see how the main difference between the classical and the quantum quantities lies in the presence of a negative term linear in $\hat n_\alpha$ in \Eq{eq: quantum <I1 Ij>}. Its origin is a direct consequence of the  photon-number quantization via the relation $\bra{n}\aop^{\dagger 2}\aop^2\ket{n}=n(n-1)$.
This immediately shows that a \emph{necessary} condition to observe a violation of the bound in \Eq{res: classical bound} is that the effective photon-number statistics has to be sub-Poissonian for some source, i.e.,
\begin{equation}\label{eq: subPoissonian}
\exists \alpha: \quad \av{\hat n_\alpha^2}_q - \av{\hat n_\alpha}^2_q \leq \av{\hat n_\alpha}_q.
\end{equation}
For example, as the squeezed vacuum is super-Poissonian, this is immediately excluded from violating \Eq{res: classical bound}, despite being considered nonclassical in other situations.
The condition in \Eq{eq: subPoissonian}, however, is not sufficient to guarantee values of $\overline G^{(Q)}$ smaller than the classical threshold, because other inputs might have large intensity fluctuations which prevent the bound from being violated. On the other hand, similarly to the two-mode case \cite{Ou_interference}, a single sub-Poissonian source could be sufficient to violate \Eq{res: classical bound}, for example when coherent states with the same average intensity as the tested input are injected in all other ports.
As could be expected, in the particular case in which all quantum sources emit coherent states (with phases randomly chosen) of amplitudes $\gamma_\alpha$, \Eq{eq: quantum <I1 Ij>} reduces to \Eq{eq: <Ii Ij>}, with $\gamma_\alpha$ playing the role of $A_\alpha$. 
 
After having shown that the violation of the bound in \Eq{res: classical bound} is possible for certain nonclassical states of light, we now study to what extent this threshold could be beaten. The presence in \Eq{eq: quantum <I1 Ij>} of a term that is linear in the number of photons makes the minimization of $\overline G^{(Q)}$ considerably harder than its classical counterpart. However, an analytical minimum can be found at least in the symmetric case where states with the same sub-Poissonian photon-number statistics $q(n)$ are injected in every input port (situation labeled by ``$sym$''). Although not general, this case is of interest in the study of many-particle interference effects, where the presence of vacuum inputs is not required as it would be in boson sampling. A symmetric setup allows a more intuitive and balanced picture and has been the study of several investigations (see Refs. \cite{Tichy2010,Lim2005,Rome_experiment}).
In this case, all the information on the input statistics is given by 
\begin{equation}\label{def: lambda}
0\leq \eta(q) = -\frac{\av{\hat n^2}_q-\av{\hat n}^2_q-\av{\hat n}_q}{\av{\hat n}^2_q}\leq 1,
\end{equation}
whose positivity signals sub-Poissonian input states while its maximum value of $1$ is reached for single-photon sources.
With an approach analogous to the classical case, we can define the complex vectors $\tilde{\psi}_i \in \mathbb C^{N=m}$ with components $\tilde{\psi}_i(\alpha) = U_{i\alpha}^*\sqrt{\mathcal E \av{\hat n_\alpha}_q/\big\langle\hat I_i\big\rangle}$.
The output correlations measured by $\overline G$ become then
\begin{equation}
\overline G^{(Q)}_{sym} = 1 + \frac{1+\eta(q)}{\binom{M}{2}}\sum_{i< j}^M \Big[|\tilde\psi_i^*\cdot\tilde\psi_j|^2 - \sum_{\alpha=1}^N |\tilde\psi_i(\alpha)\tilde{\psi}_j(\alpha)|^2\Big],
\end{equation}
where we exploited the normalization $\tilde{\psi}_i^*\cdot\tilde{\psi}_j = \delta_{i,j}$ due to the symmetry condition and the unitarity of $U$.
A comparison with \Eq{eq: class expr} immediately yields
\begin{equation}\label{eq: quantum sym min}
\min \overline G^{(Q,sym)}_{N=m,M} = 1-\frac{1+\eta(q)}{M} \leq \min \overline{G}^{(cl)}_{N=m,M},
\end{equation}
with the minimum value reached by the same optimal interferometer of the classical case, obtained by choosing $U$ to be the $m \times m$ FTM previously defined. This setup can be built in waveguides by using a number of components that scales efficiently with the system dimensionality \cite{QFTmat} and was recently proposed as a tool to distinguish real bosonic interference from semiclassical imitations, a problem of interest for the certification of boson sampling \cite{Tichy_QFT}. Our findings now show that it can also be adopted to verify the impossibility of obtaining the quantum results by means of a classical wavelike model of light.
We also note that increasing the system dimensionality reduces the allowed nonclassical range of $\overline G$ accessible by symmetric input states, as can be easily observed by comparing \Eq{eq: quantum sym min} with \Eq{res: classical bound}.

While the classical bound is completely general, the requirement of injecting states in every input port of a quantum interferometer assumes a lossless evolution. However, the same values for $\overline G^{(Q)}$ would be obtained in the presence of balanced losses, defined as independent of the path taken by the light in the interferometer. Indeed, their only effect would be the multiplication of the output intensities by a constant efficiency factor, which does not affect the studied correlation function because it simplifies in \Eq{def: Gbar}.
Balanced losses can be expected to arise with good approximation if the interferometer is symmetrically built (e.g., see the universal model recently proposed in Ref. \cite{OxfDecomposition}).

\emph{Interferometer divisibility.---}
We show how \Eq{eq: quantum sym min} allows us to develop a sufficient criterion that can certify the impossibility of dividing an interferometer into independent subblocks, therefore certifying ``true'' $m$-modes interference.
Let us consider an interferometer with outputs completely monitored, whose evolution matrix $U$ can be split into two independent submatrices.
If $m$ states of light characterized by $\eta\geq 0$ are injected into its input ports, the minimum value achievable by $\overline G^{(Q)}$ is obtained when the two subblocks are FTM matrices. Equation \eqref{eq: quantum sym min}, and the observation that $\big\langle \hat I_i \hat  I_j\big\rangle = \big\langle \hat I_i\big\rangle  \big\langle \hat I_j\big\rangle$ if outputs $i,j$ are taken in different blocks, allow us to write the aforementioned minimum as
\begin{equation}\label{res: divisibility threshold}
\min \,\overline G^{(Q,sym,div)}_{N=m,M=m} = 1 - (1+\eta)\frac{m-2}{m(m-1)}.
\end{equation}
As $\min \,\overline G^{(Q,sym,div)}_{N=m,M=m}$ is strictly larger than the global minimum of \Eq{eq: quantum sym min} with $M=m\geq 2$, a value of $\overline G$ smaller than this threshold will provide the desired certification.

\emph{Conclusions.---}
In this Letter we showed how a normalized quantifier of correlations among pairs of output intensities can yield information on the nonclassicality of the input sources and on the structure of the used multiport interferometer. In particular, we found a tight lower bound for the correlations obtained with a classical setup, where electric fields fully describe the light of the sources. We also discussed the necessity of using sub-Poissonian quantum sources in order to violate this threshold, and characterized the maximal extent of this violation under symmetric input conditions. 
Our classical bounds confirm the importance of low-order correlation functions in the study of many-particle interference effects. By comparing quantum predictions with classical electromagnetic theory, our results give a new perspective on this fundamental issue and can be of interest for experimentalists as possible tools for characterizing their setups.

\medskip
\emph{Acknowledgements.---}
We would like to thank W. S. Kolthammer, A. D. K. Plato, and A. Buchleitner for valuable discussions. We acknowledge financial support from the UK EPSRC (EP/K034480/1), the ERC grant MOQUACINO and the People Programme (Marie Curie Actions) of the EU's Seventh Framework Programme (FP7/2007-2013).

\setcounter{equation}{0}
\renewcommand{\theequation}{A\arabic{equation}}

\medskip
\section{Appendix: Proof for the classical bound}
Here we provide a proof for \Eq{res: classical bound}. First we prove that the right-hand side is a lower bound, and then that it can be saturated. To do so, it is convenient to formally write the vectors $\psi_i$ in Dirac notation: $\psi_i(\alpha) = \braket{\alpha}{\psi_i}$. Let then $H = \sum_i^M \ketbra{\psi_i}$, and note that the two following inequalities hold:
\begin{equation}\label{ineq}
	\sum_{\alpha=1}^N \bra{\alpha}H\ket{\alpha}^2\leq \Tr{H^2};\quad \Tr{H^2} \geq \frac{\Tr{H}^2}{\min\{M,N\}}.
\end{equation}
The first can be obtained by using the decomposition of the identity operator $\Id = \sum_\alpha^N \ketbra{\alpha}$ and the properties of the trace. The second follows from the inequality $\Tr{\sigma^2} \geq 1/\text{rank}(\sigma)$, which applies to any density matrix $\sigma$ because the quantum state with minimal purity is the completely mixed one, by substituting $\sigma = H/\Tr{H}$ and by noticing that rank$(H) \leq\min\{M,N\}$. Let us now rewrite \Eq{eq: class expr} as
\begin{align}
\overline G^{(cl} &= 1 + \frac{1}{M(M-1)}\left[\Tr{H^2} - \sum_\alpha^N\bra{\alpha}H\ket{\alpha}^2 \right.\\\notag
&\left.- M + M\sum_{\alpha=1}^N \sum_{i=1}^M\frac{1}{M} \Big(\bra{\alpha}\big[\ketbra{\psi_i}\big]\ket{\alpha}\Big)^2 \right].
\end{align}
The convexity of the square function allows us to lower bound the last term between square brackets with $\sum_{\alpha=1}^N \bra{\alpha}H\ket{\alpha}^2/M$. At this stage, the application of the two inequalities given in \Eq{ineq} (in the order in which they appear) leads to the desired lower bound. Its tightness can be easily proven by considering
\begin{equation}
\braket{\alpha}{\psi_i} =
\begin{cases}
\frac{1}{\sqrt{\min\{M,N\}}} \omega_M^{(i-1)(\alpha-1)} & \text{ if } \alpha\leq M \\
0  & \text{ if } \alpha > M.
\end{cases}\end{equation}


\begin{thebibliography}{51}
\bibitem [{foo()}]{footnote1}%
\BibitemOpen
\href@noop {} {\ }\bibinfo {note} {For the sake of completeness, the
	effective probability distribution $q_\alpha(n_\alpha)$ is related to
	$p_{\alpha}(\xi_\alpha)$ through the relation $q_\alpha(n_\alpha) =
	\sum_{\xi_\alpha} p_{\alpha}(\xi_\alpha) n_\alpha!
	|\varphi_\alpha^{(\xi_\alpha)}(n_\alpha)|^2$}\BibitemShut {NoStop}%
\end{thebibliography}

%

\pagebreak
\onecolumngrid
\begin{center}
	\textbf{\large Supplemental Material: Nonclassicality Criteria in Multiport Interferometry}
\end{center}
\setcounter{equation}{0}
\setcounter{figure}{0}
\setcounter{table}{0}
\setcounter{page}{1}
\makeatletter
\renewcommand{\theequation}{S\arabic{equation}}
\renewcommand{\thefigure}{S\arabic{figure}}
\renewcommand{\bibnumfmt}[1]{[S#1]}
\renewcommand{\citenumfont}[1]{S#1}

\section{Optimal Conditions in the Classical Framework}

Here we prove the optimality of fields emitted: (i) in a source- and realization-independent mode of light, (ii) with $|A_\alpha(\xi_\alpha)|$ independent of $\xi_\alpha$, and (iii) evolving with path-independent delays $\tau_{i\alpha}\equiv\tau$, for the task of minimizing the quantifier $\overline G^{(cl)}$.

To do so, it will be convenient to describe the electromagnetic field of a light pulse in Dirac notation as $ \vec{E}_{\alpha,\xi_\alpha}(t) = \langle t |\phi_\alpha^{(\xi_\alpha)}\rangle$. The vector $|\phi_\alpha^{(\xi_\alpha)}\rangle$ is defined in a Hilbert space $\mathcal H$ and can be written in the polarization-frequency domain as $\langle \omega,\lambda | \phi_\alpha^{(\xi_\alpha)}\rangle = A_\alpha(\xi_\alpha) \sqrt{\hbar \omega} \, g_{\omega,\lambda}(\xi_\alpha)$. This is consistent with Eq.(3) of the main text because
\begin{equation}
\braket{t}{\omega,\lambda} = \frac{1}{\sqrt{2\pi}} e^{-i \omega t} \vec\epsilon_{\omega,\lambda}, \qquad \sum_{\lambda}\int_{-\infty}^{+\infty} \text{d}\omega \, \ketbra{\omega,\lambda} = \Id.
\end{equation}
The norm $I_{\alpha,\xi_\alpha} = \big\langle\phi_\alpha^{(\xi_\alpha)}\big|\phi_\alpha^{(\xi_\alpha)}\big\rangle$ represents the intensity emitted by the $\alpha$-th source in the realization $\xi_\alpha$, and can be explicitly written as $I_{\alpha,\xi_\alpha} = |A_\alpha(\xi_\alpha)|^2 \mathcal E(\xi_\alpha)$, where 
\begin{equation}\label{eq: italic E}
\mathcal E(\xi_\alpha) = \sum_{\lambda}\int \text{d}\omega \, \hbar \omega |g_{\omega,\lambda}(\xi_\alpha)|^2
\end{equation}
is the energy associated with the light mode.
As the phase of $A_\alpha(\xi_\alpha)$ is chosen randomly with each realization, the average output intensity $\av{I_i}$ can be simply written as $\sum_{\alpha=1}^N T_{i\alpha} \langle I_\alpha\rangle$. 
The use of Dirac notation for the electric fields helps in rewriting the integrals appearing in the interference term of $\av{I_i I_j}$, namely
\begin{equation}
\av{\int\limits_{-\infty}^{+\infty} \text{d}t \, \vec{E}_{\beta,\xi_\beta}^*(t-\tau_{i\beta})\cdot \vec{E}_{\alpha,\xi_\alpha}(t-\tau_{i\alpha})
	\int\limits_{-\infty}^{+\infty} \text{d}t^\prime \, \vec{E}_{\alpha,\xi_\alpha}^*(t^\prime-\tau_{j\alpha})\cdot \vec{E}_{\beta,\xi_\beta}(t^\prime-\tau_{j\beta})
},
\end{equation}
in a form that is more suited to be studied with a linear algebraic approach:
\begin{equation}\label{eq: interference structure}
\sum_{\xi_\alpha,\xi_\beta}p_\alpha(\xi_\alpha)p_\beta(\xi_\beta) \bra{\phi_\beta^{(\xi_\beta)}} O^{(i)}_{\beta\alpha}\ket{\phi_\alpha^{(\xi_\alpha)}} \bra{\phi_\alpha^{(\xi_\alpha)}} O^{(j)}_{\alpha\beta}\ket{\phi_\beta^{(\xi_\beta)}} = \TrS{\mathcal H}{\Phi_\alpha O^{(j)}_{\alpha\beta}\Phi_\beta O^{(i)}_{\beta\alpha}}\av{I_\alpha} \av{I_\beta}.
\end{equation}
The left-hand side is obtained by introducing the time-shift operator $O^{(j)}_{\alpha\beta}$, diagonal in the polarization-frequency domain:
\begin{equation}
O^{(j)}_{\alpha\beta} = \int\limits_{-\infty}^{+\infty} \text{d}t \ketbraOD{t-\tau_{j\alpha}}{t-\tau_{j\beta}} = \sum_{\lambda}\int\limits_{-\infty}^{+\infty} \text{d}\omega \ketbra{\omega,\lambda} e^{-i\omega\tau_{j\alpha} +i\omega\tau_{j\beta}},
\end{equation}
while the equality in \Eq{eq: interference structure} relies on the properties of the trace and on the definition of the following Hermitian semi-positive definite operator $\Phi_\alpha$ with $\TrS{\mathcal H}{\Phi_\alpha}=1$:
\begin{equation}\label{app: def Phialpha}
\Phi_\alpha = \frac{\sum_{\xi_\alpha}p_\alpha(\xi_\alpha) \ketbra{\phi_\alpha^{(\xi_\alpha)}} }{\sum_{\xi_\alpha}p_\alpha(\xi_\alpha) \braket{\phi_\alpha^{(\xi_\alpha)}}{\phi_\alpha^{(\xi_\alpha)}}} = \frac{1}{\av{I_\alpha}}\sum_{\xi_\alpha}p_\alpha(\xi_\alpha) \ketbra{\phi_\alpha^{(\xi_\alpha)}},
\end{equation}
which characterizes the average pulse of light emitted by the $\alpha$-th source. This formalism allows us to write the correlator $\av{I_i I_j}$ in the compact form
\begin{equation}\label{eq: full classical <Ii Ij>}
\av{I_i I_j} = \av{I_i}\av{I_j} + \sum_{\alpha = 1}^{N} |T_{i\alpha}|^2 |T_{j\alpha}|^2 \left[\av{I_\alpha^2} - \av{I_\alpha}^2\right] + \sum_{\alpha\neq \beta}^{N} T_{i\alpha}T_{i\beta}^* T_{j\beta}T_{j\alpha}^* \,\TrS{\mathcal H}{\Phi_\alpha O^{(j)}_{\alpha\beta}\Phi_\beta O^{(i)}_{\beta\alpha}}\av{I_\alpha} \av{I_\beta}.
\end{equation}
The effect of the trace in \Eq{eq: interference structure} is to reduce the interference ability of the pulses of light coming from sources $\alpha$ and $\beta$ once they reach detectors $i,j$. Non-uniform evolution times are relevant because, for example,  wave-packets identical at the sources might end up shifted at the detection stage. This effect vanishes when $\tau_{i\alpha} \equiv \tau$, which implies $O^{(j)}_{\alpha\beta} \equiv \Id$. In this simpler scenario, sources $\alpha$ and $\beta$ maximally interfere when $\TrS{\mathcal H}{\Phi_\alpha \Phi_\beta}=1$, i.e., when $|\phi_\alpha^{(\xi_\alpha)}\rangle$ and $|\phi_\beta^{(\xi_\beta)}\rangle$ are all the same up to a normalization factor. Equivalently, all realizations of the two sources are characterized by the same coefficients $\{g_{\omega,\lambda}(\xi_0)\}_{\omega,\lambda}$. If $\mathcal E$ is the energy associated with this light-mode, we can substitute $\av{I_\alpha}$ with $\mathcal E \big\langle|A_\alpha|^2\big\rangle$ in all previous equations, thus retrieving the expressions given in the main text. 

The second term in \Eq{eq: full classical <Ii Ij>} is always positive and can therefore be dropped when looking for the minimum of $\overline G^{(cl)}$. The remaining expression for $\overline G^{(cl)}$ is linear in each $\Phi_\alpha$, so the minimum has to be reached when $\Phi_\alpha$ is actually the projector $\ketbra{\chi_\alpha}$ onto a pure state. These two requirements correspond to having every source emitting always the same pulse of light, potentially different from one source to the other. In this case we say that the sources are ``non-fluctuating'', and we label this condition as ``$\text{nf}$''. Let us now introduce the vectors $\ket{\psi_i} \in \mathbb C^N$ with components $\braket{\alpha}{\psi_i} = T_{i\alpha}^* \sqrt{\av{I_\alpha}/\av{I_i}}$, which generalize those defined in the main text to the case where $\mathcal E(\xi_\alpha)$ depends on the source or on the specific realization of the emitted pulse. The whole correlation function for non-fluctuating sources can then be rewritten as
\begin{equation}\label{app: intermediate eq}
\overline G^{(cl,\text{nf})}_{N,M} = 1 + \frac{1}{M(M-1)}\sum_{i\neq j}^M\sum_{\alpha\neq\beta }^{N}
\TrS{\mathcal H \otimes\mathbb C^N}{
	\ketbra{\chi_\alpha,\alpha} \tilde{\Phi}^{(j)} \ketbra{\chi_\beta,\beta} \tilde{\Phi}^{(i)}
},
\end{equation}
where $\ket{\chi_\alpha,\alpha} = \ket{\chi_\alpha}_\mathcal{H}\otimes\ket{\alpha}_{\mathbb C^N}$ and $\tilde{\Phi}^{(j)}$ is an operator acting in $\mathcal H \otimes\mathbb C^N$ defined by adsorbing the phases $\{e^{-i\omega\tau_{j\alpha}}\}_\alpha$ into $\omega$-dependent vectors $\ket{\Psi_j(w)}\in \mathbb C^N$:
\begin{equation}
\tilde{\Phi}^{(j)} = \sum_{\lambda}\int_{-\infty}^{+\infty} \text{d}\omega \, \ketbra{\omega,\lambda}\otimes \ketbra{\Psi_j(\omega)}, \qquad\text{with} \qquad \braket{\alpha}{\Psi_j(\omega)} = e^{-i\omega\tau_{j\alpha}}\braket{\alpha}{\psi_j}.
\end{equation}
Because of $\braket{\alpha}{\beta}=\delta_{\alpha\beta}$, the $N$ projectors $\ketbra{\chi_\alpha,\alpha}$ are orthogonal and have rank $1$. Therefore, the trace actually acts on a space whose dimension is effectively still $N$, with basis simply given by $\{\ket{a_\alpha}\}_{\alpha=1}^N$ where $\ket{a_\alpha} = \ket{\chi_\alpha,\alpha}$. \Eq{app: intermediate eq} can therefore be rewritten as
\begin{equation}\label{SM eq}
\overline G^{(cl,\text{nf})}_{N,M} = 1 + \frac{1}{M(M-1)}\sum_{i\neq j}^M\sum_{\alpha\neq \beta }^{N}
\Tr{
	\ketbra{a_\alpha} \tilde{\Phi}^{(j)} \ketbra{a_\beta} \tilde{\Phi}^{(i)}
},
\end{equation}
where $\tilde{\Phi}^{(j)}$ can be interpreted as a mixed density matrix in the aforementioned subspace of $\mathcal H\otimes C^N$, as can be easily verified using the completeness of $\ket{\omega,\lambda}$, the normalization $\braket{\chi_\alpha}{\chi_\alpha}=1$, and the fact that $|\braket{\alpha}{\Psi_j(w)}|^2$ does not depend on $\omega$:
\begin{equation}
\Tr{\tilde{\Phi}^{(j)}} = \sum_{\alpha=1}^N \bra{a_\alpha} \tilde{\Phi}^{(j)} \ket{a_\alpha} = \sum_{\alpha=1}^N \bra{\chi_\alpha,\alpha}\tilde{\Phi}^{(j)}\ket{\chi_\alpha,\alpha} =\sum_{\alpha=1}^{N} |\braket{\alpha}{\psi_j}|^2 \sum_{\lambda}\int_{-\infty}^{+\infty} \text{d}\omega \, |\braket{\omega,\lambda}{\chi_\alpha}|^2 = \Tr{\ketbra{\psi_j}} = 1.
\end{equation}
Therefore, a linearity argument, similar to the one used to substitute $\{\Phi_\alpha\}_\alpha$ with $\{\ketbra{\chi_\alpha}\}_\alpha$, can be adopted to state that the minimum has to be reached for operators $\tilde{\Phi}^{(j)}$ which are projectors $\big|\psi_j^\prime\big\rangle\big\langle\psi_j^\prime\big|$ onto pure states defined on the span of $\{\ket{a_\alpha}\}_\alpha$. We can now compare the expression for $\overline G^{(cl,\text{nf})}_{N,M}$ found by enforcing this condition on \Eq{SM eq} with the one that could be obtained by assuming the hypothesis (i), (ii), and (iii) since the beginning, as done in the main text [condition (ii) is taken when dropping the last positive term in Eq.(13)].
Once we write the second one in terms of the complex vectors $\ket{\psi_i}$, the two expressions are identical up to the substitution $\ket{\psi^{\prime}} \leftrightarrow \ket{\psi}$. Hypothesis (i), (ii), and (iii) thus allows the quantifier $\overline G^{(cl)}$ to reach its minimum over all independent stochastic sources and all linear evolutions.

\section{General expression for intensity correlations in the quantum framework}
In this section we want to provide general expressions for the quantities $\big\langle \hat I_i \big \rangle$ and $\big\langle \hat I_i \hat I_i \big \rangle$ in the quantum framework. When all sources emit photons in the same light-mode, the obtained result will reduce to Eq.(16) of the main text.

For a fixed realization $\vec\xi$, the phase-averaged emitted quantum state is
\begin{equation}
\hat\rho_{\vec{\xi}} = \bigotimes_{\alpha=1}^N \hat \rho_{\alpha,\xi_\alpha}= \bigotimes_{\alpha=1}^N \sum_{n_\alpha=0}^\infty |\varphi_\alpha^{(\xi_\alpha)}(n_\alpha)|^2 (\aop_{\alpha;\xi_\alpha}^\dagger)^{n_\alpha}\ketbra{0}(\aop_{\alpha;\xi_\alpha})^{n_\alpha}.
\end{equation}
In order to obtain $\big\langle \hat I_i \big \rangle$ and $\big\langle \hat I_i \hat I_i \big \rangle$, we can at first evaluate them on $\rho_{\vec\xi}$ and then take their average over the probability $p(\vec{\xi}) = \prod_{\alpha=1}^{N} p_\alpha(\xi_\alpha)$. For the first step we use the commutation relation $[\aop_{\alpha;\omega,\lambda},(\aop_{\alpha;\xi_\alpha}^\dagger)^{n_\alpha}] = n_\alpha\, g_{\omega,\lambda}(\xi_\alpha) (\aop_{\alpha;\xi_\alpha}^\dagger)^{n_\alpha-1}$, thus obtaining after some algebraic manipulations $\big\langle \hat I_i \big\rangle|_{\vec{\xi}} = \sum_{\alpha=1}^{m} |U_{i\alpha}|^2 \mathcal E(\xi_\alpha) \Tr{\hat\rho_{\alpha,\xi_\alpha} \hat n_{\alpha;\xi_\alpha}}$ and
\begin{align}
\big\langle \hat I_i \hat I_j \big\rangle|_{\vec{\xi}} &= \sum_{\alpha=1}^{m} |U_{i\alpha}|^2 |U_{j\alpha}|^2 \mathcal E^2(\xi_\alpha) \Tr{\hat\rho_{\alpha,\xi_\alpha} (\hat{n}_{\alpha,\xi_\alpha}^2 - \hat{n}_{\alpha,\xi_\alpha})} + 
\sum_{\alpha\neq \beta}  |U_{i\alpha}|^2 |U_{j\alpha}|^2 \mathcal E(\xi_\alpha) \mathcal E(\xi_\beta) \Tr{\hat\rho_{\alpha,\xi_\alpha} \hat n_{\alpha;\xi_\alpha}} \Tr{\hat\rho_{\beta,\xi_\beta} \hat n_{\beta;\xi_\beta}}\notag \\
&+ \sum_{\alpha\neq \beta}  U_{i\alpha}U_{i\beta}^*U_{j\beta}U_{j\alpha}^* \left|\sum_{\lambda}\int \text{d}\omega \, \hbar \omega \, g_{\omega,\lambda}^*(\xi_\alpha) g_{\omega,\lambda}(\xi_\beta)\right|^2
\Tr{\hat\rho_{\alpha,\xi_\alpha} \hat n_{\alpha;\xi_\alpha}} \Tr{\hat\rho_{\beta,\xi_\beta} \hat n_{\beta;\xi_\beta}},
\end{align}
where $\hat{n}_{\alpha,\xi_\alpha} = \aop^\dagger_{\alpha,\xi_\alpha} \aop_{\alpha,\xi_\alpha}$ and $\mathcal E(\xi_\alpha)$ is defined as in \Eq{eq: italic E}. 
When the average over $p(\vec{\xi})$ is performed, we are left with $\big\langle \hat I_i \big\rangle = \sum_{\alpha=1}^{m} |U_{i\alpha}|^2 \left\langle\mathcal E(\xi_\alpha) \Tr{\hat\rho_{\alpha,\xi_\alpha} \hat n_{\alpha;\xi_\alpha}}\right\rangle$ and:
\begin{align}
\big\langle \hat I_i \hat I_j \big\rangle &= \big\langle \hat I_i \big\rangle\big\langle \hat I_j \big\rangle  + 
\sum_{\alpha=1}^{m} |U_{i\alpha}|^2 |U_{j\alpha}|^2 \left\{\left\langle \mathcal E^2(\xi_\alpha) \Tr{\hat\rho_{\alpha,\xi_\alpha} (\hat{n}_{\alpha,\xi_\alpha}^2 - \hat{n}_{\alpha,\xi_\alpha})}
\right\rangle
-\left\langle \mathcal E(\xi_\alpha) \Tr{\hat\rho_{\alpha,\xi_\alpha} \hat{n}_{\alpha,\xi_\alpha}} \right\rangle^2 \right\}
\notag \\
& + \sum_{\alpha\neq \beta}  U_{i\alpha}U_{i\beta}^*U_{j\beta}U_{j\alpha}^* \left\langle\left|\sum_{\lambda}\int \text{d}\omega \, \hbar \omega \, g_{\omega,\lambda}^*(\xi_\alpha) g_{\omega,\lambda}(\xi_\beta)\right|^2
\Tr{\hat\rho_{\alpha,\xi_\alpha} \hat n_{\alpha;\xi_\alpha}} \Tr{\hat\rho_{\beta,\xi_\beta} \hat n_{\beta;\xi_\beta}}\right\rangle.
\end{align}
If the light-mode of the fields is the same for all sources and all realizations, the intensity operator $\hat I_i$ can effectively be substituted by the total number of photons in the $i$-th spatial mode multiplied by $\mathcal E$, as stated in the main text.

\end{document}